\documentclass[twocolumn,showpacs, showkeys,preprintnumbers,amsmath,amssymb, prb]{revtex4}
\usepackage{graphicx}
\usepackage{bm}
\usepackage{color}

\begin{document}

\title{Lattice and electronic anomalies of CaFe$_2$As$_2$ studied by Raman spectroscopy }

\author{K.-Y. Choi}
\affiliation{Department of Physics, Chung-Ang University, 221 Huksuk-Dong, Dongjak-Gu, Seoul 156-756, Republic of Korea}

\author{D. Wulferding and P. Lemmens}
\affiliation{Institute for Condensed Matter Physics,
 TU Braunschweig, D-38106 Braunschweig, Germany}

\author{N. Ni, S. L. Bud'ko and P. C. Canfield}
\affiliation{Department of Physics and Astronomy, Iowa State University, Ames IA 50011, USA\\
Ames Laboratory, US DOE, Iowa State University, Ames IA 50011, USA}

\date{\today}

\begin{abstract}
We report inelastic light scattering experiments on CaFe$_2$As$_2$ in a temperature
range of 4 to 290~K. In in-plane polarizations  two Raman-active phonon modes are
observed at 189 and 211 cm$^{-1}$, displaying  A$_{1g}$ and B$_{1g}$ symmetries,
respectively. Upon heating through the tetragonal-to-orthorhombic transition at about
$T_S=173$~K, the B$_{1g}$ phonon undergoes a discontinuous drop of the frequency by
4~cm$^{-1}$ whereas the A$_{1g}$ phonon shows a suppression of the integrated intensity.
Their linewidth increases strongly with increasing temperature and saturates above
$T_S$.  This suggests (i) a first-order structural phase transition and (ii) a drastic
change of charge distribution within the FeAs plane through $T_S$.

\end{abstract}

\pacs{74.70.Kn, 75.30.Gw,76.60.-k}

\maketitle \narrowtext

The recently discovered iron-based superconductors R$_x$FeAsO$_{1-x}$F$_x$ (R=La,Nd,Ga,Sm)
have triggered a burst of experimental and theoretical research activities
because of the potential relevance to high-temperature superconductors.~\cite{Kamihara,Takahashi,
chen,Ren} The Fe-based and cuprates superconductors
have remarkable similarities in structural and magnetic aspects.~\cite{grant}
The undoped compound has a two-dimensional electronic structure
and a long range spin density wave (SDW) antiferromagntic order at 134~K.~\cite{Cruz}
Doping an undoped parent compound leads to a suppression of the magnetic order
while inducing superconductivity.~\cite{Luetkens}

More recently, the ternary A$_x$M$_{1-x}$Fe$_2$As$_2$ (A=K,Na; M=Ca,Sr,Ba)
compounds have shown to share  similar structural, magnetic, and superconducting
properties with the R$_x$FeAsO$_{1-x}$F$_x$.~\cite{Rotter,Ni} Although
the superconducting transition temperatures are a little lower, the ternary compounds
have an advantage in addressing intrinsic superconducting properties
owing to the lack of oxygen and the ease of growing sizable single crystals. Among the AFe$_2$As$_2$ family CaFe$_2$As$_2$ manifests the clearest first order SDW and structural phase transition,~\cite{Ni08} making it optimally suitable for investigating the interplay between lattice and spin degrees of freedom and superconductivity.

The undoped CaFe$_2$As$_2$ has the tetragonal ThCr$_2$Si$_2$-type crystal
structure (space group I4/mmm) with lattice parameters  $a=3.912(68) \AA$ and $c=11.667(45) \AA$.
Upon cooling, a structural phase transition takes place from the high temperature tetragonal to the low temperature orthorhombic phase (Fmmm) around $T_S\sim 173$~K.~\cite{Ni08}
Drastic changes in resistivity, magnetic susceptibility, and lattice parameters
as well as narrow hysteresis suggest that the structural phase transition is of first order. In contrast, the sister compounds SrFe$_2$As$_2$ and BaFe$_2$As$_2$ exhibit a gradual change in resistivity.~\cite{Yan} This is ascribed to an extreme sensitivity of the structural instability to chemical, structural perturbations, and the presence of Sn-flux. The sharpness of the structural transition implies that CaFe$_2$As$_2$ is nearly free from such impurities. Concomitant with the structural transition, a magnetic transition to a commensurate antiferromagnetic ordering is companied with a saturated Fe moment of 0.8 $\mu_B$.~\cite{Goldman} $^{75}$As NMR measurements show a discontinuous formation of the energy gap associated with the SDW instability, giving evidence for a first order magnetic transition as well.~\cite{Baek}  Inelastic neutron scattering measurements unveiled anisotropic three dimensional magnetic behavior and a substantial spin gap.~\cite{McQueeney}

In this brief report, we present Raman scattering measurements of CaFe$_2$As$_2$ single crystals. Raman-active phonon modes show an abrupt change in frequency and linewidth around $T_S$. This is consistent with a first order nature of the structural phase transition. Significantly, the 211 cm$^{-1}$ mode jumps by 4~cm$^{-1}$ through $T_S$. Since this mode involves the displacement of Fe atoms along c-axis, this is interpreted in terms of a sensitivity of the electronic change of the FeAs plane to the out-of-plane vibration. Furthermore, the strong decrease of their linewidth below $T_S$ evidences
a drastic change of the electronic density of state at $E_F$ due to the SDW instability.

Single crystals of CaFe$_2$As$_2$ were grown out of Sn flux using
high temperature solution growth technique.~\cite{Ni08}
The high quality of the studied crystals is confirmed by an extensive
characterization by means of X-ray, neutron diffraction, thermodynamic
and transport technique.~\cite{Ni08,Goldman}
For Raman measurements, a plate-like single crystal with dimensions
of $2\times 2\times 0.1\,{\rm mm}^3$ was chosen. The sample was held in vacuum of an optical cryostat, which is cooled by a closed cycle refrigerator to helium temperature. Raman scattering experiments were performed
using the three different excitation lines $\lambda= 488$~nm (ArKr$^{+}$ Laser),
$\lambda= 532$~nm (Nd:YAG solid-state Laser), and
$\lambda= 632$~nm (HeNe Laser) in a quasi-backscattering geometry.
The laser power of 5~mW was focused to a 0.1~mm diameter spot on the surface of
the single crystal. The heating of the sample did not exceed a few K.
The scattered spectra were collected by a DILOR-XY triple spectrometer and
a nitrogen cooled charge-coupled device detector with a spectral resolution of $\sim 1$ cm$^{-1}$.

\begin{figure}
\label{fig1} \centering
\includegraphics[width=8cm]{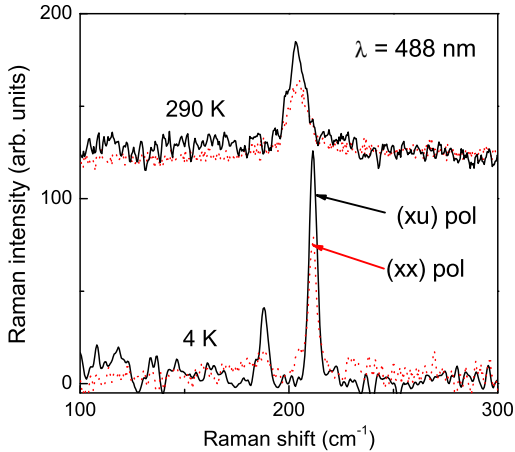}
\caption{Raman spectra in (xx) and (xu) polarization at 4 and 290~K, respectively. The solid
lines denote the (xu) polarization where the  incident light is parallel to
the $x$ axis and the scattered light is unpolarized. For the (xx) polarization, the incident
and scattered light is polarized parallel to the $x$ axis.}
\end{figure}

Figure~1 compares Raman spectra for (xx) and (xu) polarization at 4 and 290~K, respectively. In (xx) scattering geometry the incident and scattered light polarizations are parallel to the $a$ axis. In (xu) scattering configuration, the scattered light is not analyzed with respect to polarization. At room temperature we observe a single peak around 204~cm$^{-1}$. At 4~K two sharp peaks at 189 and 211~cm$^{-1}$ show up. The sharpness of the observed peaks testifies the high quality and homogeneity of the single crystal.

The factor group analysis yields four Raman-active phonon modes:
$$ \Gamma_{Raman}=A_{1g}(x^2+y^2,z^2) + B_{1g}(x^2-y^2) + 2E_{g} (xz, yz).$$
In the in-plane scattering geometry A$_{1g}$ and B$_{1g}$ modes are allowed, which correspond to the displacement of As and Fe atoms along the $c$ axis. The frequency and symmetry of the $\Gamma$-point phonons are calculated by {\it ab initio} methods
and compared to the experimental results of the isostructural compound SrFe$_2$As$_2$.~\cite{Hadjiev,Litvinchuk,Yildirim}
Thus, by referring to the comparative study in SrFe$_2$As$_2$
we are able to identify the symmetry of the two peaks. The 189 and 211~cm$^{-1}$ peaks are assigned to the A$_{1g}$ and B$_{1g}$ mode, respectively.

Here we note that the sample has a plate-like shape whose plane is perpendicular to the crystallographic c-axis. The observed Raman scattering intensity is extremely low and the sample thickness is of the order of the laser spot size.
This restricts the scattering configuration to the plane. Furthermore, we collect
the Raman spectra in the (xu) polarization because it gives a
stronger intensity than the (xx) polarization and enables us to study
the A$_{1g}$ and B$_{1g}$ modes simultaneously.

\begin{figure}
\label{fig2} \centering
\includegraphics[width=8cm]{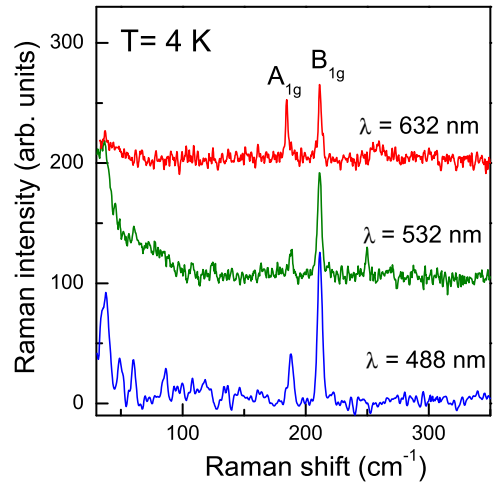}
\caption{(Online color) Comparison of Raman spectra using three different incident laser lines
$\lambda= 632, 532$, \mbox{and}\, 488\,~nm at 4~K.
}
\end{figure}

In Fig.~2 Raman spectra at three different laser lines are compared at 4~K.
All laser lines show commonly the two sharp phonon peaks at 189 and 211~cm$^{-1}$.
For the blue and green laser lines, the intensity of the 189~cm$^{-1}$ mode
is weaker than that of the 211~cm$^{-1}$ mode. The scattering intensity of both modes
becomes comparable for the red laser line. Overall, the blue and green laser lines give stronger Raman intensity than the red one. Thus, detailed temperature dependence
was measured using $\lambda= 488\,\, \mbox{and}\,\, 532$~nm.

Before proceeding, we will discuss Raman scattering on electronic excitations.
In simple metals light scattering by electrons is hardly observable at low energy
because variations of the charge density are screened by itinerant electrons,
and electrons will be collectively excited at a plasma frequency of several eV.~\cite{Devereaux}
In contrast, correlated electron systems show distinct electronic
excitations and collective modes in the optical phonon energy range. For examples, the cuprate and hydrated cobaltate Na$_x$CoO$_2\cdot y$H$_2$O superconductors show
similar flat, broad electronic continua in a certain doping range.~\cite{Peter} The
studied system exhibits no appreciable electronic excitations at least in the energy range of 40 -- 800~cm$^{-1}$ within the resolution and sensitivity of our
spectrometer. The absence of an electronic continuum might suggest that correlation effects are not as strong as the cuprate and cobaltate superconductors.
In addition, the phonon modes remain sharp and have a typical Lorentz shape rather than a Fano line shape. This implies that coupling of the phonons to other excitations is small. According to a first principle calculation,~\cite{Boeri}
electron-phonon coupling is evenly distributed among all of the phonon branches and
the electron-phonon matrix elements are extremely small due to the strongly
delocalized character of the Fe-$d$ states around the Fermi level, $E_F$.

\begin{figure}
\label{fig3} \centering
\includegraphics[width=9cm]{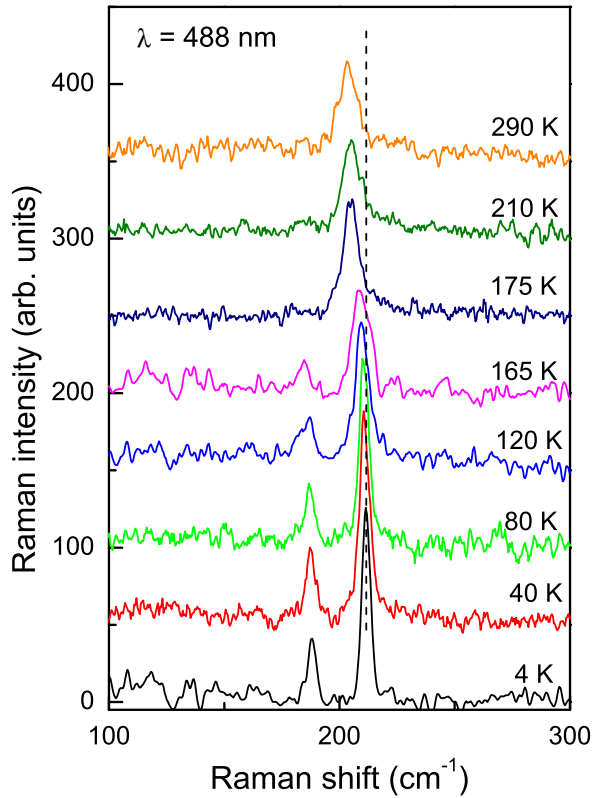}
\caption{(Online color) Temperature dependence of Raman spectra. The dashed vertical bar
indicates the position of the 211~cm$^{-1}$ mode at T=4~K.}
\end{figure}
\begin{figure}
\label{fig4} \centering
\includegraphics[width=9cm]{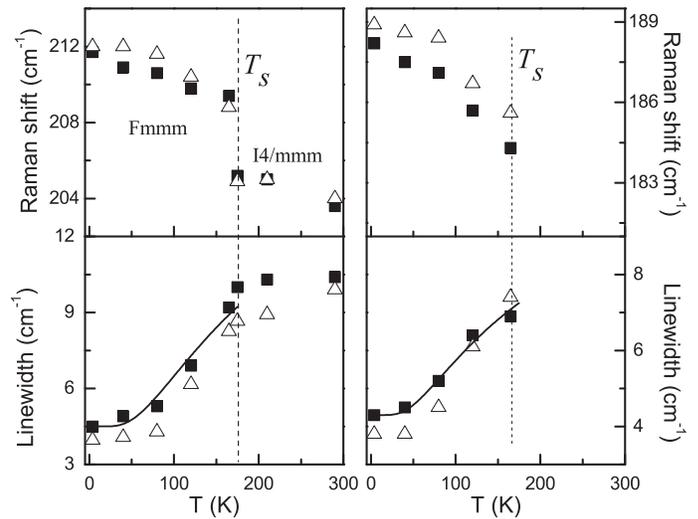}
\caption{(Left) Temperature dependence of the peak position and linewidth of
the 211~cm$^{-1}$ B$_{1g}$ mode. (Right) Temperature dependence of
the peak position and linewidth of the 189~cm$^{-1}$ A$_{1g}$ mode.
The full square represents the blue line data and the open triangle the green ones. }
\end{figure}

Figure~3 displays the temperature dependence of Raman spectra. With increasing
temperature the intensity of the A$_{1g}$ phonon is strongly reduced around $T_S$ while the frequency of the B$_{1g}$ phonon abruptly changes.
To obtain detailed information about the peak frequency and linewidth
both phonons are fitted by Lorentzian profiles.
 The results are summarized in Fig.~4. The full square
and open triangle symbols stand for the respective blue and green line data.
The errors are of the symbol size.

In the temperature region below 173~K the 211~cm$^{-1}$ B$_{1g}$ and
the 189~cm$^{-1}$ A$_{1g}$ modes soften by about 3~cm$^{-1}$ with increasing temperature.
In the narrow temperature interval around $T_S$ the 211~cm$^{-1}$ mode shows an abrupt, large jump by 4~cm$^{-1}$ and then undergoes a tiny softening by 1~cm$^{-1}$ between $T_S$ and room temperature. Its linewidth ($\Gamma$) also shows a distinct change at about $T_S$; for the 211~cm$^{-1}$ mode it increases strongly with increasing temperature and saturates above about 173~K. Below 173~K the 189~cm$^{-1}$ mode also shows a strong broadening but the saturation cannot be identified due to a suppression of the scattering intensity. The discontinuous change of the phonon frequency and linewidth at $T_S$ demonstrates a first-order nature of the structural transition as consistent with other transport, thermodynamic, and magnetic studies.~\cite{Ni08,Goldman}

We now discuss possible origins of the observed phonon anomalies.
They cannot be ascribed to the change of optical parameters since
the intensity of the 211~cm$^{-1}$ mode is largely temperature-independent (not shown here) although the 189~cm$^{-1}$ mode is strongly suppressed above T$_S$.
Next, we consider the change of interionic distances.
The 211~cm$^{-1}$ phonon frequency jumps by $2~\%$ around T$_S$.
The phonon frequency relies on a bond length, $\omega^2 \sim 1/l^3$. Between room temperature and T$_S$ the $c$-lattice parameter changes by $0.4~\%$. Thus, the lattice parameter change cannot fully explain the observed large jump of the phonon frequency. Although in-plane lattice parameters jumps by $\sim 1\%$, it will not directly couple to the B$_{1g}$ mode, which involves the out-of-plane displacements of the Fe ions. Rather, it might be related to the change of a charge distribution in the FeAs planes through T$_S$. This explains well the temperature dependence of the linewidth.

In a phonon-phonon interaction mechanism, the broadening of a phonon mode is given by
the decay of the optical phonon into acoustic modes. Since anharmonic effects are described by Boltzmann functions, the temperature dependence of a phonon linewidth does not saturate in contrast to our case.~\cite{Choi} Thus, electron-phonon interactions should be taken into account. In this mechanism, the ratio $\Gamma/\omega_0$ is given by~\cite{Axe}
$$\Gamma/\omega_0=\pi N(0)\hbar\omega_0\lambda,$$
where $N(0)$ is the electronic density of states at $E_F$, $\omega_0$ is the frequency of the specific mode, and $\lambda$ is the electron-phonon coupling parameter. As discussed above, the electron-phonon interactions are not strong due to the strong delocalization of the electronic density of state at $E_F$. However, the structural phase transition can modify
the electronic state and accordingly the strength of electron-phonon interactions can vary. Since electron-phonon interactions scale with the temperature dependence of the electronic density of sate $E_F$, the strong reduction of
the linewidth below $T_S$ can be attributed to an opening of a gap due to the SDW instability.
We find that an activated function $exp(-\Delta/k_BT)$ with $\Delta=168 - 202~K$ provides a reasonable description to the temperature dependence of the linewidth (see the solid line of Fig.~4). However, this value should not be taken literally because we cannot separate this effect from the contribution of anharmonicity to the phonon broadening. Here we mention that recent NMR measurements show a jump of an electric field gradient in the $c$ direction and an abrupt change of spin-lattice relaxation at T$_S$. This is interpreted in terms of the dramatic change of an on-site charge distribution in the As orbitals.~\cite{Baek}

To summarize, we present Raman scattering measurements of CaFe$_2$As$_2$.
The frequency and linewidth of Raman-active phonon modes show a discontinuous change
around $T_S$. This confirms a first order nature of the structural phase transition.
Significantly, the large shift of the 211 cm$^{-1}$ mode by 4~cm$^{-1}$ at $T_S$ and the strong decrease of the linewidth below $T_S$ give evidence for (i) a strong change of the electronic density of state at $E_F$ and (ii) a sensitivity of the electronic and magnetic change of the FeAs plane to the out-of-plane structure.

\begin{acknowledgments}
This work was supported by DFG and ESF-HFM. One of us (K.Y.C.)
acknowledges financial support from the Alexander-von-Humboldt Foundation.
Work at the Ames Laboratory was supported by the Department of Energy, Basic Energy Sciences under Contract No. DE-AC02-07CH11358.

\end{acknowledgments}

\end{document}